# Non-destructive detection of ions using atom-cavity collective strong coupling

Sourav Dutta* and S. A. Rangwala

*Raman Research Institute, C. V. Raman Avenue, Sadashivanagar, Bangalore 560080, India*

We present a technique, based on atoms coupled to an optical cavity, for non-destructive detection of trapped ions. We demonstrate the vacuum-Rabi splitting (VRS), arising due to the collective strong coupling of ultracold Rb atoms to a cavity, to change in presence of trapped Rb$^+$ ions. The Rb$^+$ ions are optically dark and the Rb atoms are prepared in a dark magneto-optical trap (MOT). The VRS is measured on an optically open transition of the initially dark Rb atoms. The measurement itself is fast, non-destructive and has sufficient fidelity to permit the measurement of atomic-state selective ion-atom collision rate. This demonstration illustrates a method based on atom-cavity coupling to measure two particle interactions generically and non-destructively.

## I. INTRODUCTION

Development of techniques for the detection of interaction plays a pivotal role in our understanding of the various physical phenomena in modern atomic physics. Detection of light-matter interaction, for example, has enriched our understanding of strong coupling of light and matter [1–4], electromagnetically induced transparency (EIT) [5,6], vacuum induced transparency [7] and even allowed simulation of Hamiltonians [8]. Precise measurements of matter-matter interaction, on the other hand, has enabled studies of blockade phenomena in Rydberg atoms [9], coherent association of ultracold atoms [10] and quantum chemistry of ultracold molecules [11,12]. It has also recently become possible to study well controlled hybrid systems [13–20] with different interactions. However, the detection of interaction in such systems has always been strongly perturbative and often destructive [15–20].

In this article we demonstrate in-situ detection of the ion-atom (i.e. matter-matter) interaction based on cavity mediated light-atom (i.e. light-matter) interaction. The coupling of atoms to cavities creates a versatile system for precision studies of light-matter interactions [1–3]. The system comprises two high Q oscillators, one of which is the atom and the other is a precisely tailored mode of an electromagnetic field, whose mutual interaction results in a host of fundamental phenomena [3–5,7]. Here we experimentally take this coupled system to a new regime with the addition of trapped ions in the cavity mode. In this proof of principle experiment, we demonstrate the vacuum-Rabi splitting (VRS), arising due to the collective strong coupling of the atoms to the cavity, to change in presence of trapped ions. The loss of ultracold atoms from a dark magneto-optical trap (MOT), due to ion-atom interaction, leads to a change in the atom-cavity coupling thereby allowing non-destructive detection of ions and the measurement of ion-atom collision rate.

Specifically, we use the change in collective strong coupling [1,4,21–25] of $^{85}$Rb atoms to a mode of an optical cavity to measure the interaction between ultracold $^{85}$Rb atoms in an optically dark $5s\ ^2S_{1/2}$ ($F = 2$) state and non-fluorescing $^{85}$Rb$^+$ ions that cannot be detected optically. With the cavity tuned to the non-cycling $5s\ ^2S_{1/2}$ ($F = 2$) → $5p\ ^2P_{3/2}$ ($F' = 3$) transition frequency ($\omega_{23}/2\pi$) of atomic $^{85}$Rb, we measure the VRS in two cases: with $^{85}$Rb atoms alone or with both $^{85}$Rb$^+$ ions and $^{85}$Rb atoms overlapping the cavity mode. The VRS is different in the two cases and provides a rapid, non-destructive and atomic-state selective measurement of the ion-atom collisions, which is used to determine the ion density and number of ions overlapped with the MOT.

## II. THE CONCEPT

The ion-atom interaction is facilitated in a hybrid trap [26,27], in which the ion and atom traps are precisely co-centred and aligned for optimal overlap with the cavity mode. A schematic of the experimental arrangement is shown in Fig. 1(a). The experimental concept is illustrated in Figs. 1(b)-1(d), where we show that the atom-cavity coupling results in a systematic change in the transmission of an on-axis weak probe beam depending on the presence and absence of atoms and ions. For our cavity parameters, a single atom cannot couple strongly to the cavity mode, i.e. the single atom cavity coupling constant $g_0 \ll (\gamma, \kappa)$, where $\gamma$ is the free space spontaneous decay rate of the excited atomic state and $\kappa$ is the cavity mode light loss rate due to the mirrors. Here $g_0$ is given by $g_0 = \sqrt{\mu_{23}^2 \omega_{23}/(2\hbar\epsilon_0 V_c)}$, where $\mu_{23}$ is the transition dipole moment for the $F = 2 \rightarrow F' = 3$ transition, $\epsilon_0$ is the permittivity of free space, $\hbar$ is the reduced Planck's constant and $V_c$ the cavity mode volume. When the number of atoms that occupy the cavity mode volume is increased to $N_c$, each atom can potentially couple to a single photon in the cavity mode [1,22,23]. Since any (but only one) atom at a time can couple to the photon (which is indivisible), there are $N_c$ couplings possible at each instant, which is equivalent to reducing the mode volume per atom to $V_c/N_c$. Substituting this in the expression for single atom-photon coupling gives the cavity coupling of a single photon to $N_c$ atoms as $g_N = g_0\sqrt{N_c}$ and this $\sqrt{N_c}$ amplification factor accesses the collective strong coupling regime where $g_N \gg (\kappa, \gamma)$. The present measurement is thus a straightforward amplification of the atom-photon coupling $g_0$ and shares all the features of the single atom-photon coupled system.

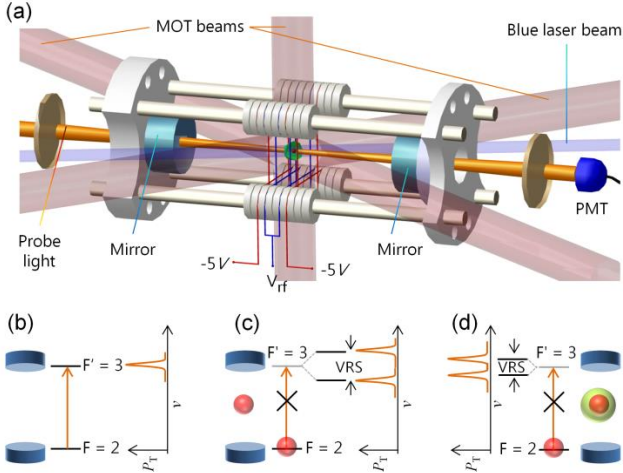

FIG. 1 (Color online). (a) Schematic of essential parts of the experiment. Ultracold $^{85}$Rb atoms in $5s\ ^2S_{1/2}$, $F = 2$ state are trapped in a dark MOT and $^{85}$Rb$^+$ ions are trapped in a spherical Paul trap such that their centres overlap. The ion-atom mixture is placed at the centre of an optical cavity and the system is probed with a weak on-axis probe beam by scanning its frequency $v$ across the $F = 2 \rightarrow F' = 3$ transition and monitoring the transmitted light using a PMT. (b) In the absence of atoms a single transmission peak is seen. (c) In presence of atoms, the regime of collective strong coupling between atoms and cavity is attained, the transmission at $\omega_{23}/2\pi$ drops to zero and two vacuum-Rabi peaks, separated in frequency by $2g_N (= 2g_0\sqrt{N_c})$, appear on either side of $\omega_{23}/2\pi$. (d) The vacuum-Rabi splitting (VRS) reduces in the presence of trapped ions.

The collective strong coupling of the atom-cavity system alters the cavity transmission as shown schematically in the Fig. 1(c), where the frequency split in the transmission through the atom-cavity system is $\Delta v = 2g_N = 2g_0\sqrt{N_c}$. When ions are co-trapped with atoms, the presence of the trapped ions alters the atom numbers in the cavity mode volume, resulting in a reduced VRS $\Delta v'' = 2g_N'' = 2g_0\sqrt{N_c''}$, as seen in Fig. 1(d). We use this change in VRS to determine the ion-atom collision rate and the density of trapped ions.

We note that the ion-atom collision rate can be obtained by measuring the changes in fluorescence [28,29] of atoms if a bright MOT is used instead of a dark MOT. However, the present VRS technique is more general and advantageous because (*a*) it measures a frequency difference rather than a power/intensity change, (*b*) the measured value of collision rate in the dark MOT is atomic quantum state specific (shown here for $F = 2$ state of $^{85}$Rb) rather than for a mixture of quantum states (e.g. mixture of $F = 2$ and $F' = 3$ state if a $^{85}$Rb bright MOT is used) and (*c*) the measurement is performed *continuously* as opposed to all previous VRS work that required turning off the MOT (e.g. [25]), thus rendering continuous measurements intractable. Further, the VRS method can be straightforwardly extended to atoms trapped in a far-off-resonance dipole trap and collectively coupled to the cavity – this can allow better state preparation, state manipulation and would get rid of near resonant light – all of which can make the method more sensitive. In addition, since measuring VRS does not destroy the atomic cloud (like absorption imaging would), it lends itself to repeated measurements on the same atomic sample.

### III. EXPERIMENTAL PROCEDURE

As shown schematically in Fig. 1(a), ultracold $^{85}$Rb atoms in the ground state $|g\rangle = |5s_{1/2}, F = 2\rangle$ are trapped in a dark MOT at a typical temperature of ~ 100-150 μK and overlapped with the mode centre of a Fabry-Perot cavity of finesse ~ 650. A modified spherical Paul trap [26,27], concentric with the MOT and cavity center, is used to trap $^{85}$Rb$^+$ ions at a typical temperature of ~ 700 K. The details of the dark MOT, the cavity and the ion trap are described in Appendix A, B and C, respectively. The empty cavity is tuned to the $|g\rangle \rightarrow |e\rangle = |5p_{3/2}, F' = 3\rangle$ transition and the frequency $v$ of the probe beam is scanned across the same transition while the transmitted power $P_T$ is monitored on a photomultiplier tube (PMT). When ultracold atoms are present in the cavity mode vacuum-Rabi peaks separated by $\Delta v\ (= 60.4 \pm 1.4$ MHz$)$ are seen (Fig. 2(a)). The VRS decreases to $\Delta v'$ in presence of the blue (473 nm) laser beam that is used to create $^{85}$Rb$^+$ ions by photo-ionization of ultracold $^{85}$Rb atoms (Fig. 2(b)). Here all the atom loss results from photo-ionization and none from ion-atom collision. Turning on the ion trap decreases the VRS further to $\Delta\nu''$ (Fig. 2(c)) due to interaction of trapped $^{85}$Rb$^+$ ions with the ultracold $^{85}$Rb atoms. Figure 2(d) shows the variation of $\Delta\nu'/\Delta\nu$ and $\Delta\nu''/\Delta\nu$ with the blue laser power. We use the measured VRS to determine, based on a rate equation model described below, the ion-atom collision rate and the local density of trapped ions.

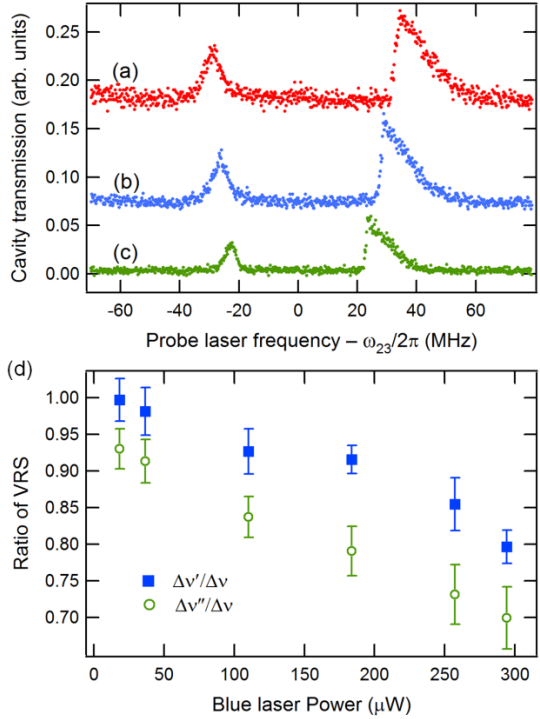

FIG. 2 (Color online). (a) VRS $\Delta\nu$ measured in presence of trapped atoms only. (b) VRS $\Delta\nu'$ in presence of trapped atoms and blue photo-ionization laser. (c) VRS $\Delta\nu''$ in presence of trapped atoms, blue laser and trapped ions. The asymmetry in the shape of the peaks is due to slight onset of optical bi-stability [24,30] but the effect is small and predominantly systematic. The small systematic errors cancel out and therefore do not affect the ensuing analysis significantly. (d) The ratios of VRS $\Delta\nu'/\Delta\nu$ (squares) and VRS $\Delta\nu''/\Delta\nu$ (circles) decrease monotonically with increase in blue laser power. The error bars represent 1 standard deviation from the mean for a set of five measurements.

## IV. ION-ATOM COLLISION RATE

The velocity of ions in the ion trap (hundreds of m/s) is much higher than the velocity of neutral atoms (which are essentially at rest). Since the depth of the ion trap (0.32 eV) is much higher than the typical depth (<1 meV) of the MOT, even a glancing elastic collision between a neutral Rb atom and a Rb$^+$ ion is always expected to lead to the neutral Rb atom being lost from the MOT. The number of atoms $N_a(t)$ in the MOT can be expressed as [28,31]

$$\frac{dN_a}{dt} = L_a - \gamma_b N_a - \gamma_p N_a - k_{aa}\int n_a^2 d^3r - k_{ia}\int n_a n_i d^3r \quad (1)$$

where $L_a$ is the atom loading rate, $\gamma_b$ is the one-body loss rate corresponding to collision with background gases, $\gamma_p$ is loss rate corresponding to the blue-laser induced photo-ionization of atoms, $k_{aa}$ is the two-body loss rate coefficient corresponding to collisions among atoms in the MOT, $k_{ia}$ is the two-body loss rate coefficient corresponding to collisions between trapped ions and atoms, and $n_a$ ($n_i$) is the density of trapped atoms (ions). The atomic density in our experiment is low and the typical loss mechanisms in a bright MOT (such as fine structure changing collisions and radiative escape [32]) are almost absent in a dark MOT [31], which allow us to neglect the $k_{aa}\int n_a^2 d^3r$ term in equation (1).

When photo-ionization and trapped ions are absent, all but the first two terms on the right hand side (r.h.s.) of equation (1) can be neglected and then solved to get the number of atoms in the dark MOT: $N_a(t) = N_a^\infty(1-e^{-\gamma_b t})$, where $N_a^\infty = L_a/\gamma_b$ is the atom number at steady state. From a fit to the experimental measurements of the faint fluorescence of the dark MOT, we obtain $N_a^\infty = 1.30(3)\times 10^6$, $L_a = 4.3(2)\times 10^5$ atoms/s and $\gamma_b = 0.331(14)$ s$^{-1}$. The MOT has a Gaussian density profile with FWHM $d \sim 400$ μm and volume $V_a \sim 5\times 10^{-11}$ m$^{-3}$ which is maintained approximately constant for all experiments. The number of atoms coupled to the cavity mode is $N_c (\approx \eta N_a^\infty)$ and leads to a VRS of $\Delta\nu = 2g_0\sqrt{N_c}$, which we directly measure using a weak probe beam along the cavity axis (Fig. 2(a)). Here $\eta$ is the fraction of the total number of atoms coupled to the cavity mode in steady state and is determined to be ~ 0.08 by comparing $N_a^\infty$ with $N_c$.

In presence of the photo-ionization laser, while the ion trap is off, the first three terms on r.h.s. of equation (1) need to be considered. The number of atoms in the MOT is then given by $N_a(t) = N_{a,p}^\infty(1-e^{-(\gamma_b+\gamma_p)t})$, where $N_{a,p}^\infty = L_a/(\gamma_b+\gamma_p)$ is the new steady state atom number. The VRS in this case is reduced (see Fig. 2(b)) to $\Delta\nu' = 2g_0\sqrt{N_c'}$, where $N_c' \approx \eta N_{a,p}^\infty$. When the photo-ionization laser and ion trap are both kept on, the last term in Eq. (1)

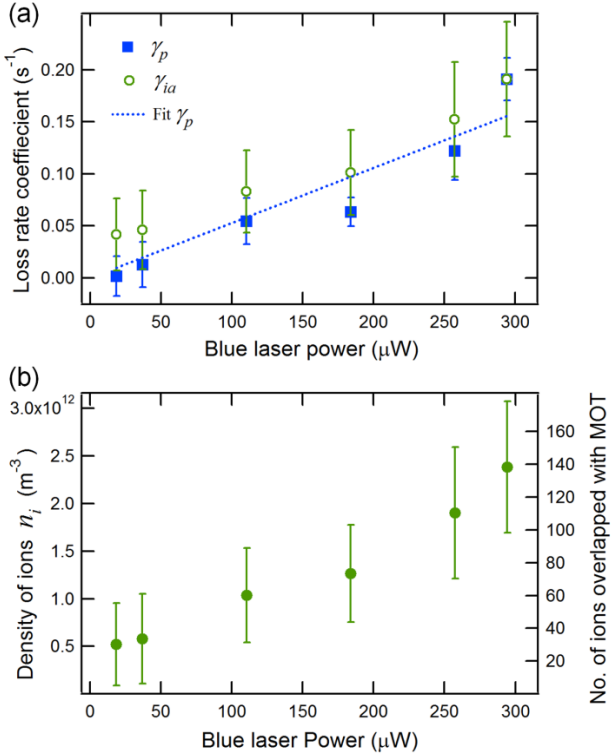

FIG. 3 (Color online). (a) The photo-ionization loss rate $\gamma_p$ (squares) and the ion-atom collision rate $\gamma_{ia}$ (open circles) plotted for different powers ($I_p$) of the blue photo-ionization laser. For the lowest powers, $\gamma_p \to 0$ within the accuracy of our measurement but $\gamma_{ia}$ is still finite because even for the smallest photo-ionization laser powers, there can be a substantial number of trapped ions overlapped with the MOT, in steady state. In addition the duration for which ions and atoms interact increases when ions are trapped. (b) The steady state ion density (left axis) at the centre of the ion trap and the number (right axis) of trapped ions within the $1/e$ radius of the MOT, determined using VRS measurements. The error bars represent the calculated errors obtained by propagation of the errors shown in Fig. 2.

also contributes. In our experiment, the trapped ions are distributed in a volume $V_i$ (~ $6 \times 10^{-8}$ m$^3$) that is much larger than the volume $V_a$ of the MOT and thus the density $n_i$ of trapped ions is, to a very good approximation, constant over the volume $V_a$ of the MOT. This allows the simplification $k_{ia} \int n_a n_i d^3 r \approx k_{ia} n_i N_a = \gamma_{ia} N_a$, where $\gamma_{ia} = k_{ia} n_i$ is the ion-atom collision loss rate. For this case, the atom number is $N_a(t) = N_{a,T}^\infty (1 - e^{-(\gamma_b + \gamma_p + \gamma_{ia})t})$, where $N_{a,T}^\infty = L_a / (\gamma_b + \gamma_p + \gamma_{ia})$ is the new steady state atom number. The VRS is further reduced (see Fig. 2(c)) to

$\Delta v'' = 2 g_0 \sqrt{N_c''}$, where $N_c'' \approx \eta N_{a,T}^\infty$. In Fig. 2(d) we plot the ratio to $\Delta v' / \Delta v$ (squares) and $\Delta v'' / \Delta v$ (circles) for different powers $I_p$ of the photo-ionization laser beam – both ratios are seen to decrease with increasing $I_p$ due to the increase in the production of ions. Using the equations above, we obtain $\gamma_p \approx \left[ (\Delta v / \Delta v')^2 - 1 \right] \gamma_b$ and $\gamma_{ia} \approx \left[ (\Delta v / \Delta v'')^2 - 1 - (\gamma_p / \gamma_b) \right] \gamma_b$. In Fig. 3(a), we plot $\gamma_p$ (squares) and $\gamma_{ia}$ (circles) for different values of $I_p$. The observed VRS, along with the proposed model, thus allows non-destructive detection of ions and the measurement ion-atom collision rate $\gamma_{ia}$. Within the range of $I_p$ values used in the experiment, we observe that both $\gamma_p$ and $\gamma_{ia}$ depend linearly on $I_p$. However, $\gamma_{ia}$ is expected to vary non-linearly with $I_p$ for higher values of $I_p$ and eventually expected to saturate. The exact form of the non-linear behavior has no bearing on the determination of ion density discussed below.

## V. DENSITY OF TRAPPED IONS

The density $n_i$ of trapped ions at the centre of the ion trap in steady state can be determined non-destructively from the relation $\gamma_{ia} = k_{ia} n_i = \langle \sigma v \rangle n_i$, where $\langle \sigma v \rangle$ is the velocity averaged product of ion-atom collision cross section $\sigma$ and the velocity of ions v. As discussed in Appendix D, $k_{ia} = \langle \sigma v \rangle \approx 8 \times 10^{-14}$ m$^3$/s. We then compute $n_i = \gamma_{ia} / k_{ia}$ and plot $n_i$ for different values of $I_p$ in Fig. 3(b) (left axis). Since the volume of the MOT, $V_a$, remains same for all values of $I_p$, we also plot the number of ions $N_{i,MOT} (= n_i V_a)$ that are overlapped with the MOT in Fig. 3(b) (right axis). The VRS measurement thus provides a non-destructive measure for the local ion density $n_i$ in the MOT region. Such a local density measurement is difficult to perform for optically dark ions such as $^{85}$Rb$^+$. For example, the other detection method [17,20] for such ions requires the extraction of all the ions in the ion trap onto a charge sensitive detector (such as a channel electron multiplier), hence losing the ions in the process. Yet another detection method [12] requires a coulomb crystal of laser cooled ions in order to sympathetically cool the dark ions.

## VI. CONCLUSION

We demonstrate an atomic-state specific and continuous method, based on collective strong coupling of atoms to a cavity, for detection of ion-atom interaction and noninvasive measurement of density of trapped ions. An important benefit of using to $N_c$ atoms, as opposed to a single atom, is that the measurements can be done much faster owing to the strong coupling being achieved even for a low finesse (i.e. high $\kappa$) cavity. The rate at which a VRS measurement can be made is fundamentally limited by the rate $\kappa$ at which photons can be extracted from the cavity and a reliable measurement requires at least $1/\kappa$ ($=\tau$) seconds. In case of a single atom, strong coupling requires $g_0 \gg \kappa,\gamma$, which in typical experiments sets the lower limit for $\tau$ to be ~ 100 μs – add to this the low duty cycle of single atom experiments. On the contrary, our experiment with $N_c$ atoms is run continuously and for a finesse ~ 650 and cavity length 4.5 cm, $\tau$ can be as low as 100 ns. The method can also be extended to detect the presence of other atoms, ions and perhaps molecules. Another interesting direction to pursue is the detection of atom-molecule conversion processes and chemistry at ultracold temperatures. We believe that the technique can be used as a generic method for in-situ, non-destructive and rapid measurement of interactions.

## ACKNOWLEDGMENTS

We thank T. Ray and S. Jyothi for the development of the experimental apparatus and M. Ibrahim for developing Fig. 1(a). S.D. acknowledges support from the Department of Science and Technology (DST), India in the form of the DST-INSPIRE Faculty Award (IFA14-PH-114).

## APPENDIX A: DARK MAGNETO-OPTICAL TRAP

The dark MOT for $^{85}$Rb is loaded from a Rb dispenser source. Six independent beams detuned by -12 MHz from the $F = 3 \rightarrow F' = 4$ transition, each 1 cm in diameter and with 7 mW power, form the cooling beams for the MOT. Two independent beams detuned by +20 MHz from the $F = 2 \rightarrow F' = 3$ transition, each 1 cm in diameter and with 2.4 mW power, form the repumping beams for the MOT. The repumping beams have their centers darkened with an opaque disc of 2 mm diameter such that no repumping light is present in the MOT region – this pumps >95% of the ultracold $^{85}$Rb atoms to the ground non-fluorescing $F = 2$ state. The low fluorescence of the trapped atoms in a dark MOT minimizes the otherwise deleterious effects of fluorescent photons being coupled to the optical cavity. Apart from the VRS measurements, we independently measure the number of atoms, density and the loading rate in the dark MOT by instantaneously turning it bright using a repumping light tuned to the $|5s_{1/2}, F = 2\rangle \rightarrow |5p_{3/2}, F' = 3\rangle$ transition and recording the fluorescence on a calibrated PMT.

## APPENDIX B: CAVITY PARAMETERS

The Fabry-Perot cavity consists of a pair of curved mirror (radius of curvature 50 mm) separated by $L = 45.7$ mm and with cavity waist $w_0 = 78$ μm. The cavity finesse is measured to be ~ 650. The single atom-cavity coupling constant is given by $g_0 = \sqrt{\mu_{23}^2 \omega_{23}/(2\hbar\epsilon_0 V_c)}$; where $\mu_{23}$ is the transition dipole moment for the $F = 2 \rightarrow F' = 3$ transition separated in energy by $\hbar\omega_{23}$, $\epsilon_0$ is the permittivity of free space and $V_c$ is the cavity mode volume. With these parameters, collective strong coupling between atoms and cavity is achieved when $N_c > 3\times10^4$ or equivalently when $N > 4\times10^5$. Here $N_c$ is the number of atoms coupled to the cavity mode and $N$ is total number of atoms in the MOT. The weak probe beam used for VRS measurements has a few nW of circulating power corresponding to an intensity of a few μW/cm$^2$ inside the cavity. This intensity is chosen so that it is high enough to give a good signal to noise ratio for single-shot VRS measurements but is still low enough that the anharmonicity in the vacuum Rabi peaks does not affect the measurements significantly.

## APPENDIX C: ION TRAPPING

The $^{85}$Rb$^+$ ions are created continuously by ionizing, with a 473 nm laser beam, a fraction of the residual excited $5p_{3/2}$ state atoms present in the dark MOT. The ions are trapped in the modified spherical Paul trap [26] consisting of four wires (Fig. 1(a)) - a sinusoidal voltage of amplitude $V_{rf} = 85$ V and frequency 500 kHz is applied to the inner pair of wires (separated by 3 mm) while both the outer wires (separated by 6 mm) are at -5 V. The secular frequencies in the radial and axial directions are measured to be $f_r = 43$ kHz and $f_z = 54$ kHz, respectively. The maximum displacement $\delta_z$ in the axial direction is ~ 2.5 mm from which the upper bound of the trap depth $U_z$ is estimated to be ~ 0.32 eV using the expression $U_z = (1/2)m_{Rb}(2\pi f_z)^2 \delta_z^2$, where $m_{Rb}$ is the mass of $^{85}$Rb$^+$.

# APPENDIX D: ESTIMATION OF COLLISION RATE COEFFICIENT

Determining the collision rate coefficient $k_{ia} = \langle \sigma v \rangle$ requires the knowledge of ion temperature and velocity. The temperature and the most probable velocity $v_m$ of the trapped $^{85}$Rb$^+$ ions are estimated assuming a Maxwell Boltzmann (MB) velocity distribution. To a very good approximation $k_{ia} = \langle \sigma v \rangle \approx \sigma_m v_m$, where $\sigma_m$ is the cross-section at $v_m$. The trap depth of 0.32 eV determines the highest velocity of a trapped ion to be 850 m/s. The MB distribution should thus have negligible fraction (say, < 2%) of ions above 850 m/s – the constraint is satisfied by a MB distribution at a temperature of ~ 700 K and $v_m = 360$ m/s. The ion-atom collision cross-section at velocity $v_m$ is calculated using the expression [33]

$$\sigma_m = \pi \left( \frac{m_{Rb} C_4^2}{2\hbar^2} \right)^{1/3} \left( 1 + \frac{\pi^2}{16} \right) \left( \frac{1}{2} m_{Rb} v_m^2 \right)^{-1/3},$$

where $C_4 = 1.09 \times 10^{-56}$ Jm$^4$ [34] determines the long range ion-atom interaction potential $V = -C_4/2R^4$. Using these we compute $k_{ia} = \langle \sigma v \rangle \approx \sigma_m v_m = 8 \times 10^{-14}$ m$^3$/s.

---